\documentclass{elsart}
\usepackage{graphicx}
\usepackage{amssymb}

\begin{document}

\begin{frontmatter}
\title{The Economic Mobility in Money Transfer Models}
\author{Ning Ding}, \author{Ning Xi} \and \author{Yougui Wang\corauthref{cor1}}

\corauth[cor1]{Corresponding author.
\\ {Tel.: +86-10-58807876; Fax:+86-10-58807876.}
\\{\em E-mail address:\/}\, ygwang@bnu.edu.cn (Y. Wang)}
\address{Department of Systems Science, School of Management, Beijing Normal University, Beijing,
100875, People's Republic of China}

\begin{abstract}
In this paper, we investigate the economic mobility in four money
transfer models which have been applied into the research on
wealth distribution. We demonstrate the mobility by recording the
time series of agents' ranks and observing their volatility. We
also compare the mobility quantitatively by employing an index,
``the per capita aggregate change in log-income'', raised by
economists. Like the shape of distribution, the character of
mobility is also decided by the trading rule in these transfer
models. It is worth noting that even though different models have
the same type of distribution, their mobility characters may be
quite different.
\end{abstract}

\begin{keyword}
Economic Mobility\sep Wealth Distribution\sep Transfer Model

\PACS 89.65.Gh \sep 87.23.Ge \sep 05.90.+m
\end{keyword}
\end{frontmatter}

\section{Introduction}\label{int}

The issue of income or wealth has been one of most popular topics
among economists all along. With the study on this issue, the
performance and wellbeing of an economy can be shown. So far, an
enormous body of work has been devoted to the two aspects of this
issue: distribution \cite{pareto,ben,roy,teulings,champ} and
mobility \cite{mob1,mob2,mob3}. The former is the information that
one could use to determine how many economic agents have each
level of income or wealth. The latter is the phenomenon that the
agents' income or wealth varies in economy over time. It is
obvious that the distribution is static by nature, while the
mobility is dynamic.

The study can be traced back more than 100 years ago. An Italian
economist, Vilfredo Pareto, declared that the income distributions
in several European countries uniformly follow Power laws
\cite{pareto}. He believed that the income distribution is
considerably stable both over time and across countries and there
must be an uniform mechanism which governs the shape and formation
of distribution. Benoit Mandelbrot observed that the distribution
of income converges to Pareto distribution for many different
definitions of income, for example income could be written as the
sum of agricultural, commercial or industrial incomes; in cash or
in kind and so on \cite{ben}. Based on the description and shape
of income distribution, the measurement methods of the inequality
or well-being were discussed \cite{sen,cowell}.

When the study went further, economists began to realize that only
analysis on the distribution is not sufficient especially when
exploring the generating mechanism of income and measuring the
degree of inequality \cite{kuznets,jj}. When we compare the degree
of inequality in two economies with the same distribution but
different dynamics, it is obvious that the static analysis based
on the data of a given moment cannot provide a complete picture,
and the dynamic analysis is needed. Consequently, the study on the
income mobility, which is one important aspect of the dynamics,
got started \cite{mob1,mob2,mob3}.

So far, the economists devoted their efforts much more to the
concept, measurement and empirical research on this issue, but a
few to the mechanism behind the distribution and mobility. Several
models were constructed to explain the shape of the distribution
by introducing the relationship between the wages and the
abilities \cite{roy,teulings,champ}. On the other topic, G. C.
Loury tried to employ the stochastic theory to research the
mobility among successive generations of workers \cite{ecnomob}.
The lack of deep investigation in this field dose not suit with
the significance of this issue, which calls for more efforts.

When physicists forayed into the economic domain with their
advanced tools, they concerned the statistical characteristics of
some economic variables, such as return in financial markets,
volatility of GDP. Recently, researchers turn to carry out some
empirical work on the income or wealth distributions and its
growth or variation \cite{emp1,emp2,emp3,emp4,souma}. Comparing
with this aspect, the more important and suitable landing zone for
physicists is the mechanism behind phenomena.

Many efforts had been put into exploring the formation mechanism
of distribution \cite{solomon,bouch,slanina}. Besides, a series of
multi-agent models had been constructed in which the interaction
between agents is simplified as random transfer of money in a
closed system \cite{redner,basic,saving1,saving2,dn}. Some of
these works are reviewed in a popular journal \cite{hayes}, and
the related models are called as money transfer models
\cite{transfer}. The basic model was built by A. Dr\u{a}gulescu
and V.M. Yakovenko basing on the analogies between money transfer
in trading and energy transfer in molecule collision \cite{basic}.
The simulation results show that the distribution of money obeys a
Boltzmann-Gibbs law. A group led by B.K. Chakrabarti introduced
the saving behavior into the model \cite{saving1,saving2}, and
found the money distribution follows a Gamma law when all the
agents are set with the same saving factor, but changes to a power
law as the saving factor is set randomly. N. Ding et al introduced
the preferential behavior into the trading process and obtained a
stationary power-law distribution \cite{dn}.

Indeed, analyzing the relationship between the distribution and
the trading rule is helpful to investigate the mechanism. In fact,
distribution is just one of the windows through which how the
economy works can be observed. In money transfer models, where the
economy is assumed to be closed, continuous transfer of money
among agents leads to stationary distribution of money. However,
agents still switch their positions or ranks even after the
distribution gets steady. Thus, in order to explore the mechanism
behind the distribution, static analysis on distribution is not
sufficient. Suppose two cases where the steady distributions are
observed to be the same, but their fluctuation modes of agents'
ranks are different. To distinguish them, the analysis on mobility
is in need, for it can provide more information as to the
mechanism behind the distribution due to its dynamic and onymous
nature.

Moreover, the study on mobility in the proposed models makes the
evaluation criteria more complete. The aim of econophysicists to
develop these models is to mimic the real economy by abstracting
its essence. We cannot judge whether such abstraction is
reasonable or not depending on the shape of distribution only. In
other words, even though the distribution resulted from the
proposed model is close to the reality, we are not convinced that
it is reasonable. The mechanism might yield a different mobility
character from the reality. Thus, we must take mobility into
account when constructing a ``right" multi-agent model. For
instance, to develop a model to simulate the Japanese income
distribution, the theoretical prediction on mobility must be
tested against the empirical results in Ref. \cite{souma}.

In addition, the transfer models proposed by econophysicists are
also available for understanding the formation of mobility.
Especially, these models embody the essential character of
mobility, for agents in the assumed economy always change their
position by exchanging the money. This phenomenon has been
referred to in Hayes's review article \cite{hayes}.

In this paper, the mobility in four typical transfer models is
observed and analyzed. In the next section, we make a brief review
of the researches of economists on the mobility by which our work
can be erected and from which the measurement index we employed
comes. In Section 3, the four models and simulations are
introduced. The results are analyzed and compared in Section 4.
The final section summarizes.

\section{Income Mobility}
Since the 1960s, economists gradually realized static snapshots of
income distribution alone is not sufficient for meaningful
evaluation of wellbeing and the income mobility must also be
weighted. Kuznets \cite{kuznets} once declared that two societies
``... may differ greatly in meaning because of different degrees
of internal mobility ...", even if they have identical size
distributions of income. And, Jarvis and Jenkins \cite{jj}
argued:`` To some people, greater inequality at a point in time is
more tolerable if accompanied by significant mobility...".

That can be understood easily from a simple example. Suppose in an
economy there are two individuals with income \$1 and \$2
initially. At the next moment, their incomes change to \$2 and \$1
respectively. The distribution in this case is unchanged, but the
rank of either individual varies over time. Although the system
seems unequal at either of the two moments in terms of the
distribution, the fact is that the two individuals are quite equal
combining these two moments. Besides, from the simple example, it
can also been found out that the structure of economy may varies
enormously with an unchanged distribution. So, the investigation
on mobility is helpful to the measurement on equality. It should
be noted that mobility is not income's particular character. There
are also mobilities in some other economic concepts, such as
wealth, consumption, labor and etc. Thus, the research on income
mobility is also helpful to the studies on those aspects.

In contrast to the meaning of mobility research, there is less
consensus on the concept and measurement. Some researchers view
mobility as a reranking phenomenon: people just switch their
position. To measure this purely relative mobility, some indexes
are used, such as the coefficient of rank correlation, the average
jump in rank and the mean number of absolute ranks changed. In the
view of other researchers, any change of individuals' income leads
to mobility. Here, it is an absolute concept. Naturally, the
income mobility can be decomposed into two parts: one induced by
reranking phenomenon and the other induced by distributional
change. To measure this absolute mobility, G. S. Fields and E. A.
Ok started from a series of axioms and chose an index whose
expression is $l=\frac{1}{N}\sum^N_{k=1}|\log x_{k0}-\log
x_{k1}|$, where $x_{k0}$ and $x_{k1}$ are the initial income and
final income of agent $k$ respectively \cite{gfield}. This index,
named as ``per capita aggregate change in log-income'', is
actually one kind of distance between two states of an economy.
The ``distant'' index not only can reflect the change of income in
aggregate level, but also is sensitive to the individuals' switch
in rank.

P. V. Kerm went further and divided the mobility induced by
distributional change into two parts: dispersion and growth
\cite{pvkerm}. Moreover, the empirical analysis in the same work
shows that income mobility are essentially due to `reranking'
mobility. In fact, the `reranking' mobility can be deemed as the
transfer of income among people with given aggregate income. By
identifying the income with money, this kind of the `reranking'
phenomenon can be demonstrated in the money transfer models where
the total amount of money is conserved. Thus the mechanism of
mobility can be analyzed based on these models that will be
introduced in the next section.

\section{Transfer Models and Simulations}

The construction of transfer models provides a powerful tool for
the study on the income or wealth distribution and mobility. The
economies assumed in these models are pure monetary systems which
are only composed of agents and money. The money is held by agents
individually. And the agents exchange money to each other
according to the trading rule which ensures the non-negativity of
any agent's money and the conservation of the total money. The
simulation results show that no matter how uniformly and
forcefully one distributes money among agents initially, the
successive trades eventually lead to a steady distribution of
money. And the shape of money distribution is only determined by
the trading rule. There exist considerable transfer models which
are used in the study on distribution. From these models, we
choose the following four typical ones to show the mobility
phenomena. Since the scale and the initial distribution have no
effect on the final result, all of the simulations of these four
models were carried out with $N=1,000$ agents and $M=100,000$
units of money. The amount of money held by each agent is set to
be $M/N=100$ at the beginning.

\subsection{Model I: Ideal Gas-like Model}

By identifying money in a closed economy with energy and the trade
between agents with two-body elastic collision in an ideal gas, A.
Dr\u{a}gulescu and V.M. Yakovenko proposed the ideal gas-like
model. In each round of this model, two agent $i,j$ are chosen
randomly to take part in the trade. As to which one is the payer
or receiver, it is also decided randomly. Supposing the amounts of
money held by agent $i$ and $j$ are $m_i$ and $m_j$ respectively,
the amount of money to be exchanged $\Delta m$ is then expressed
as follows:
\begin{equation}\label{basic}
    \Delta m=\frac{1}{2}\varepsilon (m_i+m_j)
\end{equation}
where, $\varepsilon$ is a random number from zero to unit. If the
payer cannot afford the payment, the trade is cancelled.

\subsection{Model II \& III: Uniform Saving Rate Model and Diverse Saving Rate Model}

B.K. Chakrabarti etc. thought the gap between reality and the
ideal gas-like model is too huge. Some features, such as saving
behavior, should be considered. They argued that the people always
keep some of money in hand as saving when trading. So, B.K.
Chakrabarti etc. developed the ideal gas-like model by introducing
the saving behavior.

They employed the trading pattern of ideal gas-like model that two
agents are chosen out to participate in the trading in each round
at random. The difference is that agents keep a part of money as
saving as they participate in the trade. And the ratio of saving
to all of the money held by one agent is denoted by $s$ and called
saving rate in this paper. The saving rates of all the agents are
set before the simulations. Suppose that at $t$-th round, agent
$i$ and $j$ take part in trading, so at $t+1$-th round their money
$m_i(t)$ and $m_j(t)$ change to
\begin{equation}\label{aaa}
m_i(t+1)=m_i(t)+\Delta m; m_j(t+1)=m_j(t)-\Delta m;
\end{equation}
where \[
    \Delta m=(1-s)[(\varepsilon-1)m_i(t)+\varepsilon
    m_j(t)];
\]
and $\varepsilon$ is a random fraction between zero and unit. It
can be seen that if $\Delta m$ is positive, agent $i$ is the
receiver of the trade, otherwise the payer. This model degenerates
into the ideal gas-like model if $s$ is set to be zero. In this
model, all of agents are homogenous with the same saving rate. So
we call it uniform saving rate model.

B.K. Chakrabarti etc. modified the model by introducing diverse
saving rate. They set agents' saving rates randomly before the
simulations and keep them unchanged all through the simulations.
Likewise, this model is called diverse saving rate model.
Correspondingly, the trading rule Equation (\ref{aaa}) is
transformed into
\begin{equation}\label{bbb}
m_i(t+1)=m_i(t)+\Delta m; m_j(t+1)=m_j(t)-\Delta m;
\end{equation}
where \[
\label{deltam2}
    \Delta m=(1-s_i)(\varepsilon-1)m_i(t)+(1-s_j)\varepsilon m_j(t);
\]
and $s_i$, $s_j$ are the saving rates of agent $i$ and $j$
respectively.

\subsection{Model IV: Preferential Dispensing Model}

We attribute two forces to the formation of power-law
distribution: one is preference in dispensing, which lets the rich
have higher probability to get the money; the other one is
stochastic disbursement, which ensures that the rich would be
deprived of wealth and become the poor in some day. In order to
verify this idea, we proposed a new money transfer model named as
preferential dispensing model and obtained a steady power-law
distribution of money.

In each round of this model, every agent pays money out, and the
ratio of payment to the amount of money he held is determined
randomly. To any unit of these money, a receiver is chosen
randomly from the agents except the payer. The probability
$p_{i,j}$ at which agent $i$ (potential receiver) gets the money
from agent $j$ (payer) is given by
\begin{equation}\label{ptg}
    p_{i,j}=\frac{m_i}{\sum_{n=1 (n\neq j)}^{N}m_n},
\end{equation}
where $m_i$ is the amount of money held by agent $i$ before the
trade. The constraint $n\neq j$ in this rule eliminates the
possibility for the payer to get back the money he has paid out.

\section{Results and Discussion}
To show the `reranking' mobility, at the end of each round, we
sort all of agents according to their money and record their
ranks. Here, we choose the rank rather than the amount of money to
show the mobility because the fluctuation of rank can reflect the
mobility especially the `reranking' mobility with an unchanged
distribution. Another reason is that the former is too sensitive
to the perturbation. In these transfer models, agents keep to
switch their rank over time whether the distribution is stationary
or not. To avoid the effect of transients, all of data are
recorded after the money distributions get stationary. It should
be noted that the time intervals of sampling are different for
these models as collecting the data. The data are sampled every
1000 rounds\footnote{Which is identical to the number of agents
$N$.} in the first three models and every round in model IV. We
know that only a pair of agents take part in trade in each round
of the first three models, while all of agents do so in the Model
IV. It is obvious that the mobilities are not in the same level if
the sampling intervals are the same in these models. In order to
make the mobilities comparable, we proceed sampling in the way
mentioned above, for the times per agent taking part in trading in
one round in model IV are equal to those in 1000 rounds in the
first three models. In this way, we get the time series of rank
for all agents. Some typical time series of rank from the sampling
data of these four models are shown in Figure 1 correspondingly.
The rank is inversely related to the amount of money held, so the
more money an agent held, the lower his curve is in these figures.

From the time series of ranks plotted in Figure 1, we can compare
the character of rank fluctuation of these four models with each
other. It can be seen that the fluctuations in model I and model
II are quite similar except the frequency(Fig.1a and 1b). All of
the agents can be the rich and be the poor. The rich have the
probability to be poor and the poor also may be luck to get money
to be the rich. In other words, the economies in these two models
are quite fair where every agent is equal in the opportunity. As
mentioned in Section 3, the model I is a special case of model II
when the uniform saving rate is zero. Thereby the Figure 1a and b
can be taken as the results of the cases where the uniform saving
rate are $0$ and $0.5$ respectively. Comparing them, it can be
seen that the lower the saving rate, the higher fluctuation
frequency, which means lower saving rate leads to greater mobility
in model II. From Fig.1c, we can see that the economy in model III
is highly stratified where the rich always keep their position,
and the poor are doomed to being the poor. Moreover, the agents in
different level differ in the rank fluctuation. The higher the
agent's ranks, the smaller the variance of his ranks. From Fig.1d,
each of the agents has the same probability to be the rich in the
model IV just like model I. However, the time interval that agents
keep in their current position is quite longer, which results in a
lower degree of mobility.

Now we turn to quantitative measurement of mobility. Although
there are quite a few indexes for measuring the relative mobility,
they fall into disuse due to various deficiencies. In contrast,
the measurement index raised by G. S. Fields et al has the
advantage over them. It can be applied for not only the absolute
mobility but also the relative one, as long as we take the agents'
ranks as the sampling variable. So we define the `reranking'
mobility between the two samples recorded at different moments as
follows:
\begin{equation}\label{distance}
    l(t,t')= \displaystyle
    \frac{1}{N}\sum_{i=1}^{N}|\log(x_i(t))-\log(x_i(t'))|,
\end{equation}
where $x_i(t)$ and $x_i(t')$ are the ranks of agent $i$ at $t$ and
$t'$ respectively. It is obvious that the bigger the value of $l$,
the greater the level of mobility. The values of $l$ of the four
models are shown in Table 1. These results are gotten in the
following way. Firstly, more than 9000 samples are recorded
continuously after the distribution gets steady. Secondly, the
`distance' between any two consecutive samples is reckoned.
Finally, the average of these `distances' is calculated as $l$.
Just as mentioned above, in the first three models, the difference
between $t$ and $t'$ is 1000, while that in model IV is 1.

The data given in Table 1 verify our primary conclusion about
model I and II derived from Fig.1a and 1b in some degree: the
mobility decreases when the saving rate increases. The intuition
for this result is straightforward. The more money agents keep
when trading, the less money they pick out to participate in the
trade. And so the less probability of change in rank. Therefore,
the higher saving rate, the lower mobility.

Comparing with model II, the value of mobility is quite small in
model III. This is obviously due to the stratification. However,
what surprised us is that the value is much smaller than that in
model II when the uniform saving rate is $0.5$, even than that
when the uniform saving rate is set to be $0.9$. As we know, the
agents in the economy assumed interact with each other and the
index cannot be obtained by simple mathematical averaging. As the
agents with high ranks almost do not change their positions, their
contribution to the mobility is nearly none. Although the rest of
agents move drastically, the range of their rank fluctuation is
limited. As a result, the mobility in model III is very small.
With regard to stratification, the diverse saving rates must be
culprit because they are the only denotation of the agents'
heterogeneity. To demonstrate this point, we show the relation
between the agents' ranks and their saving rates in Fig.2. It can
be seen clearly that there exists the negative relation between
the rank and the saving rate. And the rank fluctuation of agents
with low saving rate is more drastic than that of agents with high
saving rate.

Now, we turn to the mobility in Model IV. From Table 1, we can see
that no stratification appears in this case and the degree of
mobility is the smallest. The homogeneity of agents' behavior
pattern in this model gives every agent an equal chance to get
rich. However, the preferential behavior ensures the agents
keeping their position for a relatively long period. As a result,
the assumed economy evolves with very low mobility and without
stratification. Comparing model III and model IV, we can conclude
that although both of the monetary distributions follow Power
laws, their mobility characters are totally different.

From the above discussion, it can be seen that the index can help
us to compare the mobility quantitatively. However, the index
itself can not give the whole picture of mobility. The ``distant"
index we used in this paper can be rewritten as
\begin{equation}\label{distancere}
    l(t,t')= \displaystyle
    \frac{1}{N}\sum_{i=1}^{N}|\log(vol_i(t,t')+1)|,
\end{equation}
where $vol_i(t,t')$ actually denotes the volatility of agent $i$'s
rank. It takes the following form
\begin{equation}\label{vol}
vol_i(t,t')=\frac{x_i(t')-x_i(t)}{x_i(t)}.
\end{equation}
The new expression indicates that the index is a presentation of
volatility of rank at aggregate level. To examine the distribution
of volatility can provide us the information of mobility in more
detail.

The distributions of the rank volatility for the four models are
illustrated in Figure 3 respectively. It is noted that the
distributions in model I and II are quite similar and their right
tails follow power laws. In model II, we further calculated the
exponent of the power-law distribution $p(x)\varpropto x^{\alpha}$
for different uniform saving rates. As shown in Fig.4, the
exponent $\alpha$ varies with the saving rate in an exponential
way. This result is consistent with the dependence of the mobility
index on the saving rate. When the saving rate gets higher, the
money exchanged will be less and the volatility of rank will
decrease. Consequently, when the saving rate increases, the right
side of volatility distribution will shift to the vertical axis,
leading to a more steeper tail. From Figure 3c and 3d, it is clear
that the distributions of the rank volatility for models III and
IV are quite different. The tail of volatility distribution of
model III converges to an exponential line as the times of
simulations increase. In model IV, the part of the volatility
distribution within the range from 0 to about $1.5$ has a good
fitting to an exponential line, but the tail scatters no matter
how many times of the simulations are performed.

\section{Summary}

To economists, mobility is a supplement to distribution when the
equalities are compared. To physicists, it is an approach to the
mechanism behind the distribution in microscope. The purpose of
this paper is to investigate the mobility phenomena in four
typical money transfer models. Some characters of mobility are
also presented by recording the time series of agents' ranks and
observing the distribution of rank volatility. We find that the
mobility is also decided by the trading rule in these transfer
models. Our finding implies that the economies with the same
distribution may be different in mobility. It would be helpful to
take the character of mobility into account when constructing a
multi-agent model on the issue of distribution.

\section*{Acknowledgement}

We thank Prof. Zengru Di for his help discussions and comments.
The work is partially supported by NSFC under grant No.73071072.

\newpage

\begin{figure}
  \includegraphics[width=0.48\textwidth]{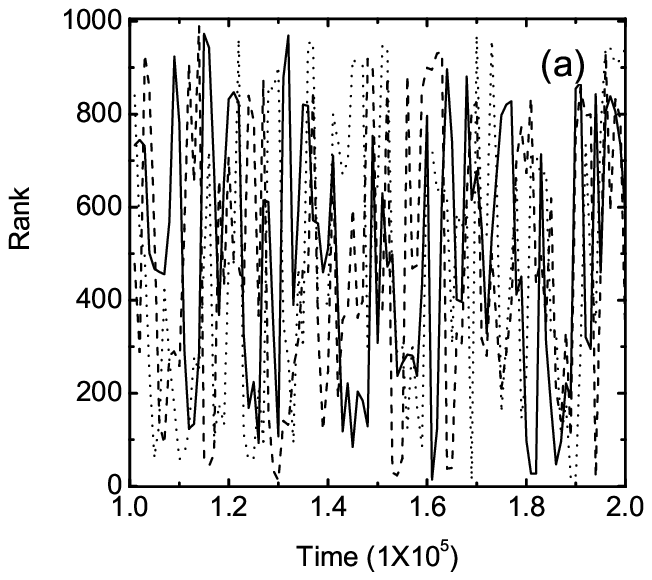}\ \
  \includegraphics[width=0.48\textwidth]{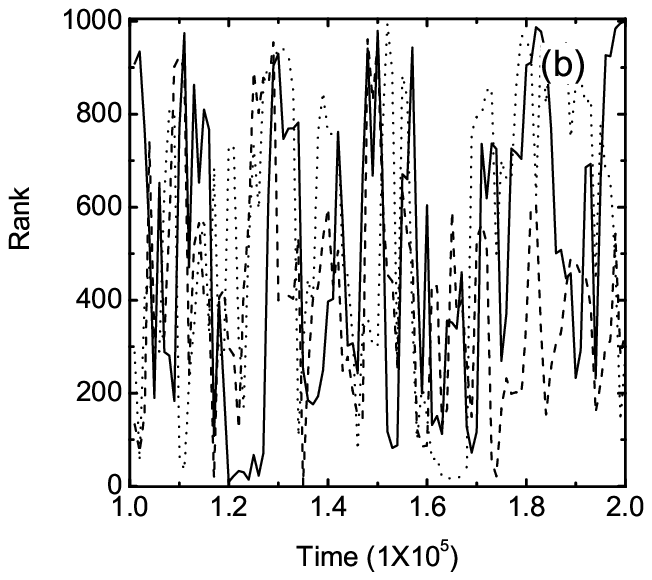}\\
  \includegraphics[width=0.48\textwidth]{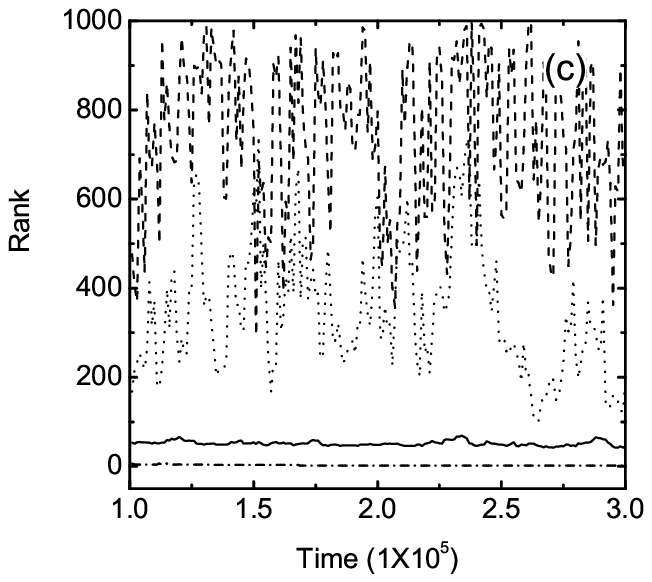}\ \
  \includegraphics[width=0.48\textwidth]{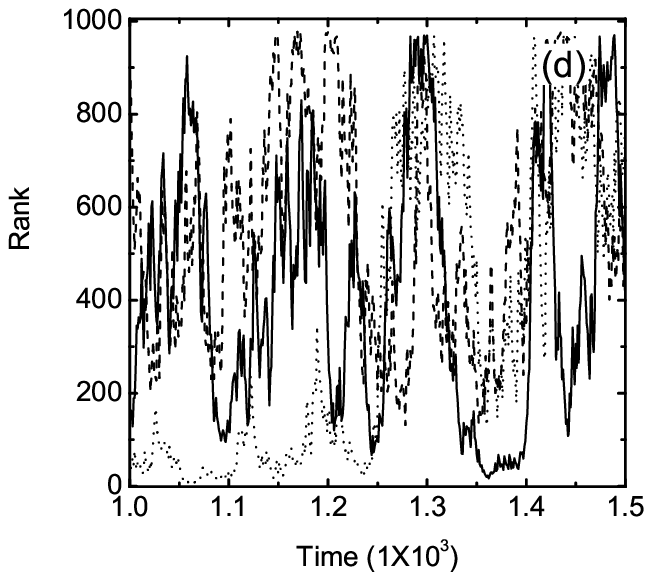}
  \caption{The typical time series of ranks from model I: Ideal Gas-like Model (a),
  model II: Uniform Saving Rate Model with saving rate $0.5$ (b),
  model III: Diverse Saving Rate Model where the saving rates of these typical
  agents are 0.99853, 0.9454, 0.71548 and 0.15798 (from bottom to top respectively) (c)
  and model IV: Preferential Dispensing Model (d).
  }\label{1}
\end{figure}

\begin{figure}
  \includegraphics[width=0.96\textwidth]{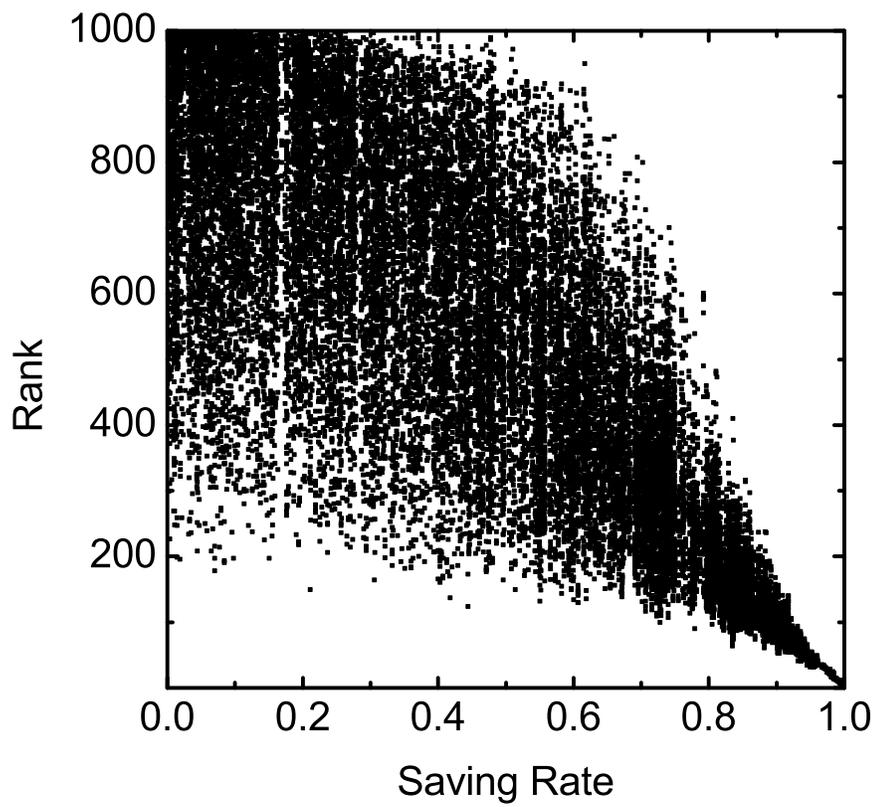}
  \caption{The correlation between agents' ranks and saving rate in model III.}\label{2}
\end{figure}

\begin{figure}
  \includegraphics[width=0.48\textwidth]{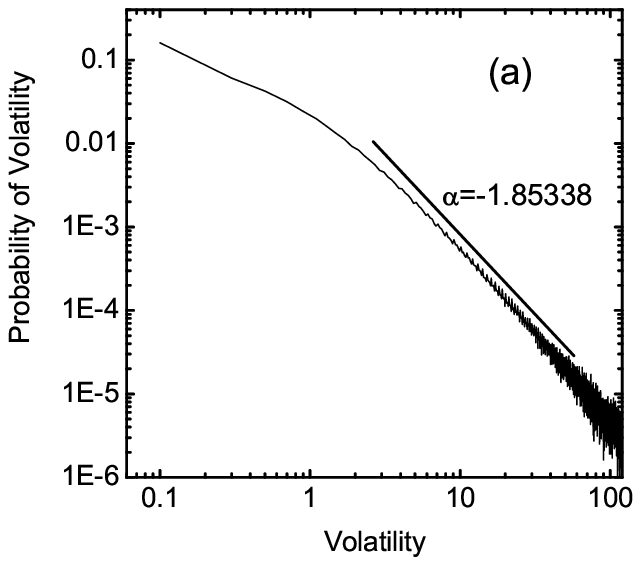}\ \
  \includegraphics[width=0.48\textwidth]{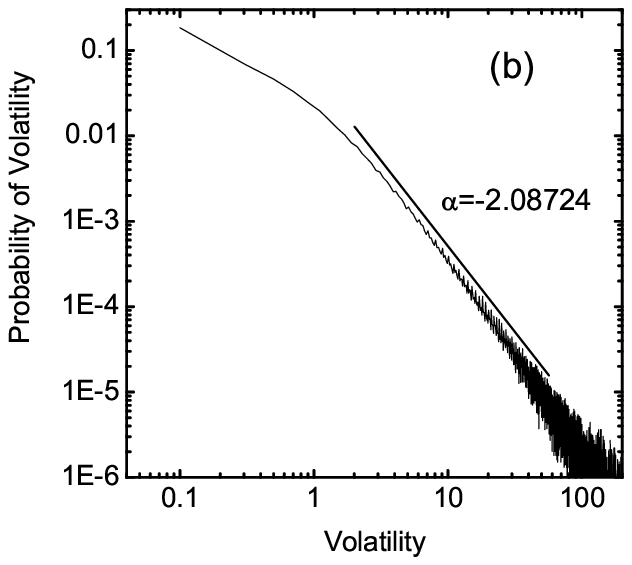}\\
  \includegraphics[width=0.48\textwidth]{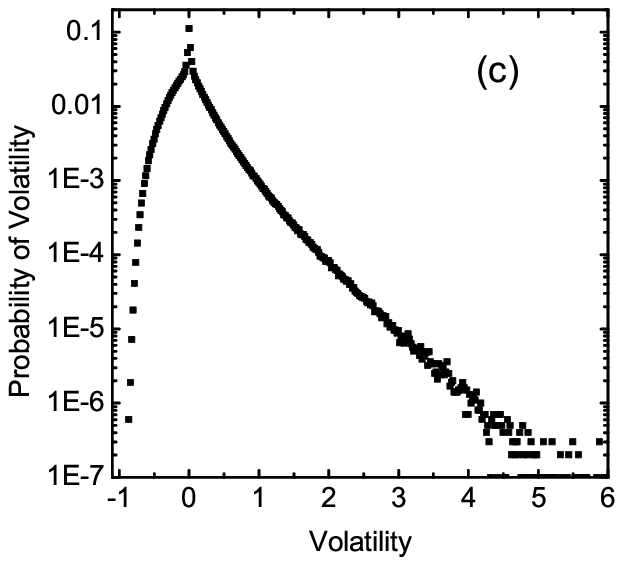}\ \
  \includegraphics[width=0.48\textwidth]{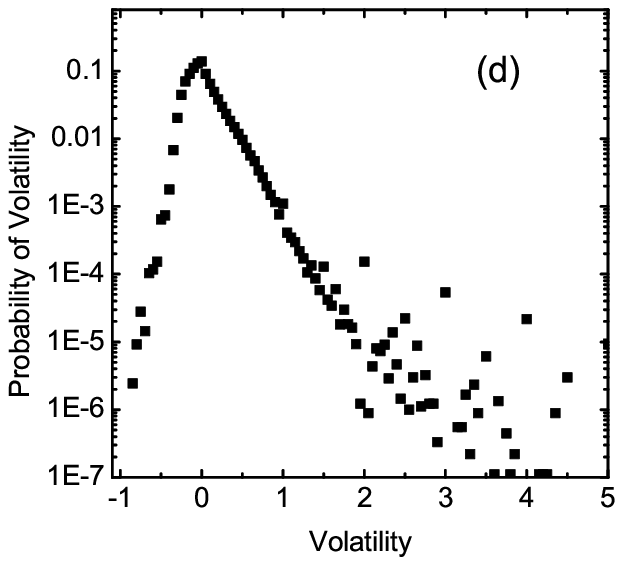}\\
  \caption{The distribution of the volatility of agents' ranks from model I: Ideal Gas-like Model (a),
  model II: Uniform Saving Rate Model with saving rate $0.5$ (b),
  model III: Diverse Saving Rate Model (c)
  and model IV: Preferential Dispensing Model (d).}\label{3}
\end{figure}

\begin{figure}
  \includegraphics[width=0.96\textwidth]{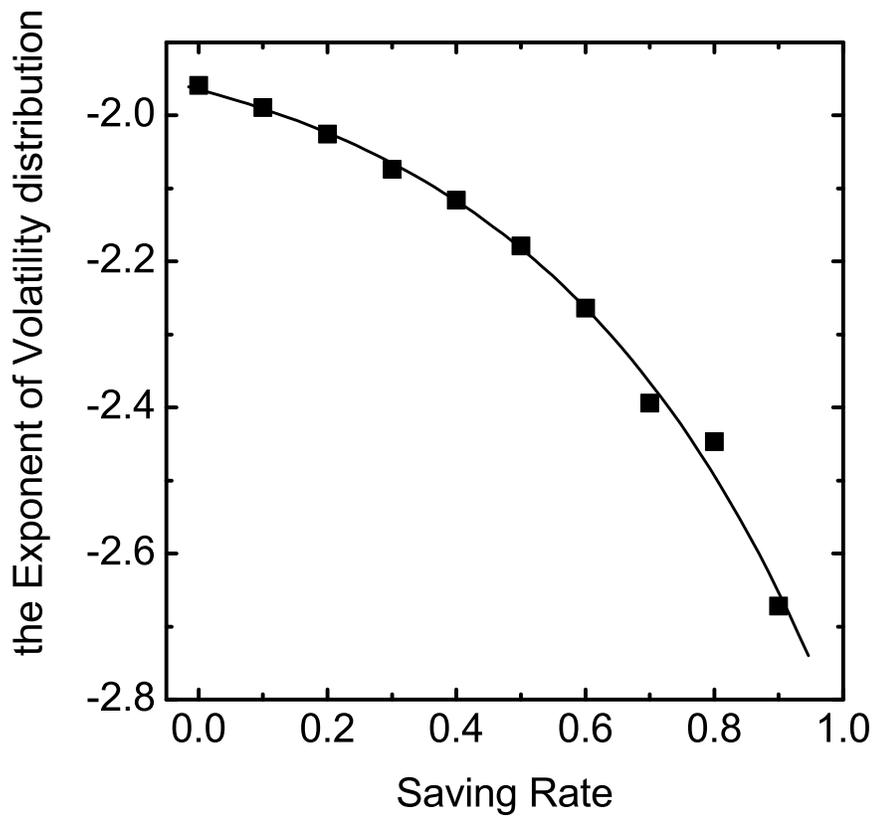}
  \caption{The relation between the uniform saving rate and the slope of the volatility distribution in model II:
  simulation results(dots) and fitness result(curve). $\alpha=-0.104e^{s/0.443}-1.860$
}\label{4}
\end{figure}


\clearpage

\begin{table}
 \caption{Comparison of Four
Typical Transfer Models With Respect to Monetary Distribution and
Mobility}\label{a}
\begin{tabular}{c|c|c|c}
  \hline
   & Monetary Distribution & Mobility $l(t,t')$ & Stratification \\
  \hline
  Model I & Exponential Law & 0.72342 & No \\
  \hline
  Model II & Gamma law &   & No \\
  $s=0.1$  & &  0.70269 & \\
  $s=0.3$  & &  0.65165 & \\
  $s=0.5$  & &  0.58129 & \\
  $s=0.7$  & &  0.4773 & \\
  $s=0.9$  & &  0.30212 & \\
  \hline
  Model III & Power Law & 0.19671 & Yes \\
  \hline
  Model IV & Power Law & 0.14828 & No \\
  \hline
\end{tabular}
\end{table}


\begin{thebibliography}{99}
\bibitem{pareto} V. Pareto, \textit{Cours d'Economie
Politique\/},  Macmillan, Paris, 1897.

\bibitem{ben} B. Mandelbrot, Int. Econ. Rev. \textbf{1}(2), May, 79 (1960).

\bibitem{roy} A. D. Roy, Oxford Economic Papers \textbf{3}(2), June, 135 (1951).

\bibitem{teulings} C. N. Teulings, J. Polit. Econ. \textbf{103}(2), April, 280 (1995).

\bibitem{champ} D. G. Champernowne, Econ. J. \textbf{LXIII}, June,
318 (1953).



\bibitem{mob1} B. R. Schiller, Am. Econ. Rev. \textbf{67}(5),
926 (1977).

\bibitem{mob2} A. F. Shorrocks, Economica \textbf{50}, 3 (1983).

\bibitem{mob3}G. S. Feilds and E. A. Ok, in \textit{ Handbook of Inequality Measurement\/},
edited by J. Silber, Kluwer Academic Publishers, Dordrecht, 2003.
p. 557-596.

\bibitem{sen} A. Sen, in \textit{ Ethical Measurement of Inequality: Some Difficulties\/},
edited by W. Krelle and A. F. Shorrocks, Nothholland Publishing
Company, Nwtherlands, 1977. p. 81-94.

\bibitem{cowell} F. A. Cowell, Rev. Econ. Stud. \textbf{XLVII}(3), April, 521 (1980).

\bibitem{kuznets} S. S. Kuznets, \textit{Modern Economic Growth: Rate, Structure and
Spread\/}, Yale University, New Haven, 1966. p.203.

\bibitem{jj} S. Jarvis and S. P. Jenkins, Econ. J. \textbf{108}, 1 (1998).

\bibitem{ecnomob} G. C. Loury, Econometrica \textbf{49}(4), 843 (1981).

\bibitem{emp1} W. Souma, Fractals \textbf{9}, 463 (2001).

\bibitem{emp2} M. Levy and S. Solomon, Physica A \textbf{242}, 90 (1997).

\bibitem{emp3}  A. Dr\u{a}gulescu and V. M. Yakovenko, Eur. Phys.
J. B \textbf{20}, 585 (2001).

\bibitem{emp4} A. Dr\u{a}gulescu, and V. M. Yakovenko, Physica A
\textbf{299}, 213 (2001).

\bibitem{souma}  Y. Fujiwara, W. Souma, H. Aoyamac, T. Kaizoji, M. Aoki, Physica A \textbf{321}, 598 (2003).

\bibitem{solomon} M. Levy and S. Solomon, Int. J. Mod. Phys. C \textbf{7}, 595 (1996).

\bibitem{bouch} J.-P. Bouchaud and M. Mezard, Physica A \textbf{282}, 536 (2000).

\bibitem{slanina} F. Slanina, Phys. Rev. E \textbf{69}, 046102 (2004).

\bibitem{redner} S. Ispolatov, P. Krapivsky, and S. Redner, Eur. Phys. J. B \textbf{2},
267 (1998).

\bibitem{basic} A. Dr\u{a}gulescu and V. M. Yakovenko, Eur. Phys.
J. B \textbf{17}, 723 (2000).

\bibitem{saving1} A. Chakraborti and B. K. Chakrabarti, Eur.
Phys. J. B \textbf{17}, 167 (2000).

\bibitem{saving2}A. Chatterjee, B. K. Chakrabarti and S. S.
Manna, Physica A \textbf{335}, 155 (2004).

\bibitem{dn} N. Ding, Y. Wang, J. Xu and N. Xi, Int. J. Mod. Phys. B \textbf{18}(17-19), 2725
(2004).

\bibitem{hayes} B. Hayes, Am. Scientist \textbf{90}, 400 (2002).

\bibitem{transfer} Y. Wang, N. Ding, N. Xi, To appear in \textit{Practical Fruits of Econophysics },
edited by H. Takayasu, Springer-Verlag, Tokyo, 2005.



\bibitem{gfield} G. S. Fields and E. Ok, Economica \textbf{66}, 455 (1999).

\bibitem{pvkerm} P. V. Kerm, Economica \textbf{71}, 223 (2004).





\end{thebibliography}
\end{document}